 \documentclass[final,3p,times,twocolumn]{elsarticle}

 \usepackage{graphicx}

\usepackage{amssymb}

\usepackage[usenames]{color}

\biboptions{comma,square}

\journal{Solid State Communications}

\begin{document}

\begin{frontmatter}



\title{Low temperature magnetic transitions of single crystal HoBi}


\author[address1,address2]{Ant\'on Fente}
\author[address1,address2]{Hermann Suderow\corref{cor1}}
\author[address1,address2]{Sebasti\'an Vieira} 
\author[address2,address3]{Norbert Marcel Nemes}
\author[address2,address3]{Mar Garc\'ia-Hern\'andez}
\author[address4]{Sergey~L. Bud'ko}
\author[address4]{and Paul~C. Canfield}

\address[address1]{
	Laboratorio de Bajas Temperaturas, Departamento de F\'isica de la Materia Condensada, Instituto de Ciencia de Materiales Nicol\'as Cabrera and Condensed Matter Physics Center (IFIMAC), Facultad de Ciencias, Universidad Aut\'onoma de Madrid, E-28049 Madrid, Spain}
		
\address[address2]{
	Unidad Asociada de Bajas Temperaturas y Altos Campos Magn\'eticos, UAM/CSIC, Cantoblanco, E-28049 Madrid, Spain}
	
\address[address3]{
	Instituto de Ciencia de Materiales de Madrid, Consejo Superior de Investigaciones Cient\'ificas,	Campus de Cantoblanco, 			E-28049 Madrid, Spain.}

\address[address4]{
	Ames Laboratory and Departament of Physics and Astronomy, Iowa State University, Ames, Iowa 50011, USA}

\cortext[cor1]{Corresponding Author.\\ \textit{Email Address:} hermann.suderow@uam.es}
	
\begin{abstract}
	We present resistivity, specific heat and magnetization measurements in high quality single crystals of HoBi, with a residual resistivity ratio of 126. We find, from the temperature and field dependence of the magnetization, an antiferromagnetic transition at 5.7 K, which evolves, under magnetic fields, into a series of up to five metamagnetic phases.
\end{abstract}

\begin{keyword}
	A. Holmium \sep A. Bismuth \sep A. Single Crystal \sep C. Magnetic Transitions

\end{keyword}

\end{frontmatter}



\section{Introduction}

Rare-earth compounds provide an exceptional frame to study magnetic interactions. 4f and 5f shells are incomplete giving often magnetic moments. Magnetic ordering is controlled by the interplay between exchange interactions and crystal field coupling \cite{Canfield06}. Some rare earth based materials are superconducting and magnetic \cite{Canfield06,Crespo09}, with phase diagrams showing a mutual relationship between metamagnetic and superconducting transitions \cite{Galvis12}. These phase diagrams have brought considerable interest \cite{Canfield98, Buzdin85, Aoki01, Petit10}.\\

To further advance, it is useful to characterize magnetism in simple metallic systems with a crystal structure which is easy
to handle. Of interest, in particular for first principles calculations, are the monopnictide compounds. They crystallize in a NaCl-like structure, what makes them suitable to perform theoretical analysis. Therefore these compounds have attracted interest for testing magnetic interaction theories \cite{Mullen74, Pawlicki88} and more recently due to their semiconducting properties and practical applications \cite{Bhardwaj11}. A structural transition into a CsCl structure appears with pressure \cite{Bhardwaj11, Svane01}. The electronic phase diagram of an entire range of rare earth monopnictides \cite{Petit10} has been predicted using \textit{ab-initio} calculations. Experiments and theory have addressed several systems also in the particular case of Ho-monopnictides, such as HoP \cite{Fischer85_1}, HoAs \cite{Schmid85} or pure and Y doped HoSb \cite{Kim76, Hessel80, Jensen80, Taub74}.\\

HoBi is a metallic compound. Its lattice parameter is of $a=0.62291(3) nm$ \cite{Pearson} and its magnetic behavior stems from the interaction between the Ho ions. Ho$^{3+}$ has total angular momentum of J=8 and the magnetic behavior is expected to be correspondingly rich. Previous studies on HoBi single crystals, grown from the melting of polycrystalline samples, find at low temperatures \cite{Ott84} an antiferromagnetic transition at 5.7 K based on thermal expansion and x-ray measurements, and a structural transition into a pseudotetragonal cubic structure with an axial ratio $c/a$ greater than unity. Neutron diffraction studies \cite{Fischer85} performed in polycrystalline samples find a second order transition into a fcc type II antiferromagnet. Recent studies use density functional theory to predict structural an electronic properties of this compound \cite{Coban10}. Bhajanker {et al} using also theoretical calculations have predicted a structural transition under pressure at 26 GPa into a CsCl-like structure for HoBi \cite{Bhajanker12}.
None of the studies we found about HoBi addresses the evolution of its magnetic phases with both field and temperature.\\

In this paper we describe the synthesis and characterization of high quality single crystals of HoBi. We study in particular temperature and magnetic field dependence of the magnetization with the aim to build a H-T phase diagram at low temperatures. We find typical features of a second order antiferromagnetic transition and a number of field-induced transitions. Values found for the effective Ho-magnetic moment and saturation magnetization agree with previously reported ones  \cite{Ott84}.\\


\section{Material and Methods}

\begin{figure}[ht!]
	\begin{center}
		\includegraphics[width=0.4\textwidth,clip]{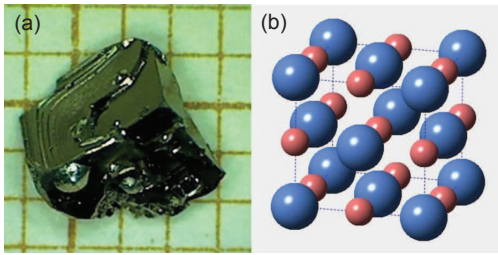}
	\end{center}
	\caption{(a)HoBi crystal grown from excess of Bi flux. Each of the small squares in the background corresponds to one millimeter. (b) Crystallographic structure of HoBi where blue (big) and red (small) spheres represent Ho and Bi ions respectively.}
	\label{crystal}
\end{figure}

High quality single crystals of HoBi were grown from Bismuth flux \cite{Growth_1, Growth_2} using a 10\% atomic percentage of Ho and 90\% of Bi. Pure elements were placed into an alumina crucible and heated from room temperature to $1000\,^{\circ}\mathrm{C}$ in three hours, cooled in another three hours to $900\,^{\circ}\mathrm{C}$ and slowly cooled in 104 hours to $400\,^{\circ}\mathrm{C}$. As a result we obtained crystals like the one in figure \ref{crystal} with sizes between 0.2 mm and 1 mm. Crystals show a cubic-like shape according to their $NaCl$ structure and are easily cleaved along the main crystallographic axes.\\ 
	
Magnetization measurements were performed in a Quantum Design Magnetic Properties Measurement System (MPMS) in zero field cooling and with the external field applied parallel to the [100] direction. M(H) curves were measuring stabilizing the temperature and going up and down in field. Since practically no hysteresis was found we only display here measurements when increasing the magnetic field. Regarding M(T) curves we stabilized the field and we went up in temperature. For resistance measurements we use a Quantum Design Physical Properties Measurement System (PPMS) and the four probe method with platinum wire contacts glued with silver epoxy. HoBi is an air sensitive material and in order to reduce exposure we did not polish or shaped the sample. This condition makes it hard to give an accurate resistivity value due to the relevance of the geometric factor in this calculation. Zero field specific heat was measured between 2 K and 20 K using adiabatic microcalorimetry again with a PPMS system.\\

\section{Results and Discussion}

\begin{figure}[ht!]
	\begin{center}
		\includegraphics[width=0.9\columnwidth,clip]{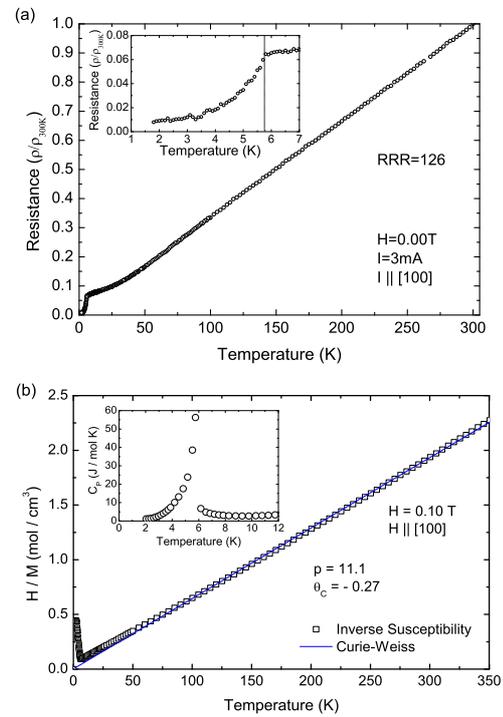}
	\end{center}
	\caption{In (a) we show the temperature dependence of the resistance normalized to its ambient temperature value at zero magnetic field. Current is applied along the a axis of the cubic structure. In (b) we show the temperature dependence of the inverse susceptibility with a magnetic field of 0.1 T applied along the a axis. Blue line is a fit to the expression given in the text between 55 K and 350 K. Inset: Zero field specific heat shows a jump at the magnetic transition.}
	\label{quality}
\end{figure}


The temperature dependence of the resistance of HoBi normalized to its ambient temperature value is shown in Fig.\ref{quality}.a from 300 K to 1.8 K. The resistance linearly drops down to 50 K, and then saturates. Below 5.7 K, at the magnetic transition temperature, we observe a significant drop. The RRR (Residual Resistance Ratio) is 126 between 300 K and 1.8 K.\\

Fig.\ref{quality}.b shows the inverse susceptibility at a magnetic field of 0.1 T. We find a linear Curie-Weiss-like behavior until HoBi enters the magnetically ordered phase at 5.7 K. This transition is consistent with the drop in resistance and therefore we can associate the latter to a significant decrease of magnetic scattering. Magnetic ordering increases the electron mean free path \cite{Canfield95}.\\

We can fit the susceptibility as a function of temperature using 
	$$ \chi = \frac{C}{T-\Theta_C}, $$
and obtain between 55 K and 350 K a Curie-Weiss temperature of $\Theta_C$=-0.26 K. The effective Bohr magneton is extracted from
	$$C=\frac{N\mu_B^2}{3Vk_B}\ p^2 ,$$
giving a value of p=11.1. The free Ho$^{3+}$ ion has a Bohr magneton per Ho of 10.6. We find a value slightly bigger than the expected one but in the same order. Below the transition the sample presents the typical temperature dependence of the susceptibility for an antiferromagnet.\\

The specific heat (inset of Fig.\ref{quality}.b) shows a clear lambda like anomaly with a peak at the magnetic transition temperature again consistent with the transition we observed in susceptibility and resistance measurements and with a dependence very similar to the one found for HoSb in ref. \cite{Taub74}.\\

\begin{figure}[hb!]
	\begin{center}
		\includegraphics[width=0.9\columnwidth]{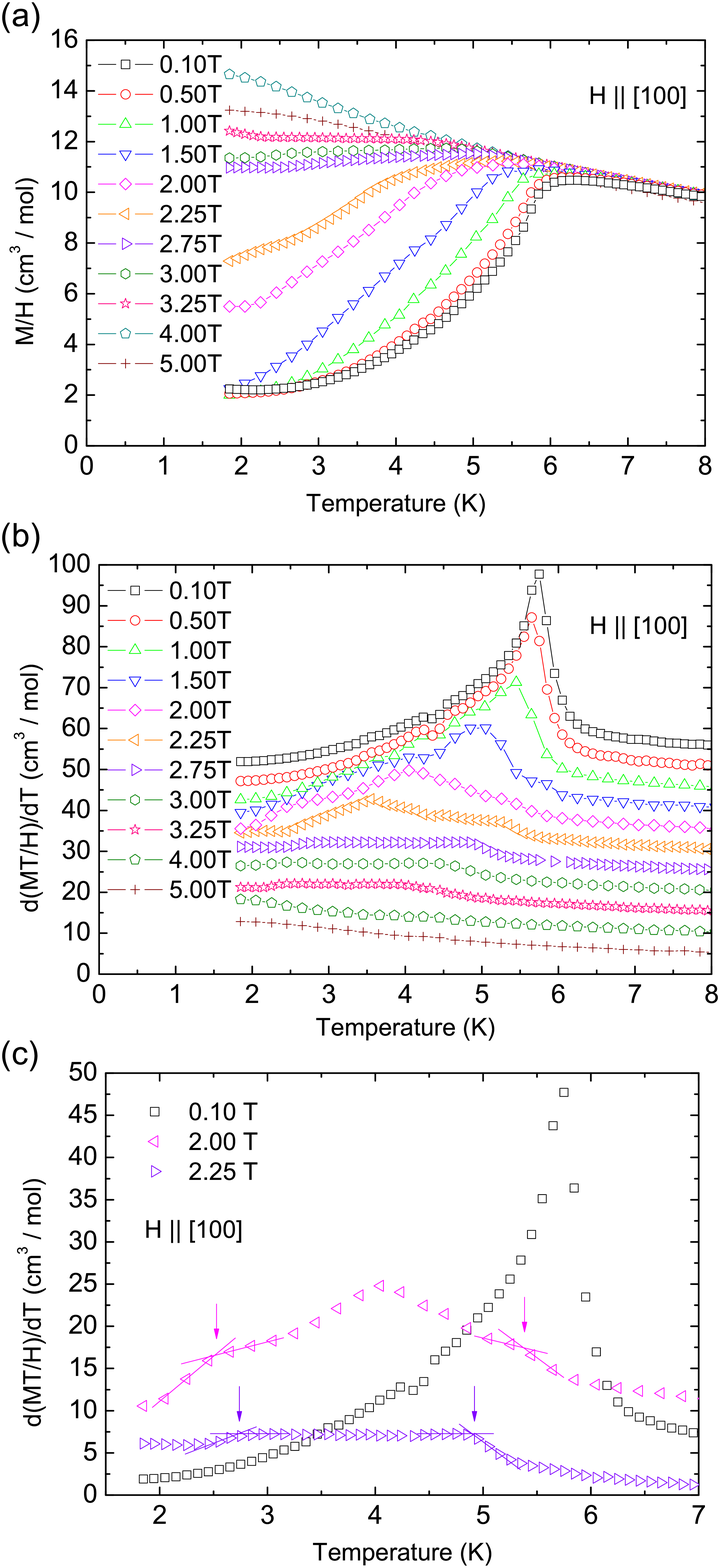}
	\end{center}
	\caption{(a) M/H for different fields show antiferromagnetic behavior below the transition temperature for the lower fields. Between H=2 T and H=4 T different metamagnetic states are reached at the lowest temperatures and for field above 4.00 T no transition is seen. (b) We show $d(MT/H)/dT$ to better highlight the transition temperatures \cite{Fisher62}. Data for different temperatures have been shifted by 0.001 cm$^3$/mol. (c) Detail of $d(MT/H)/dT$ for some fields to show the types of features found in this derivatives and the criteria used for the selection of the transition temperatures. The H=2 T and the H=2.25 T data are shifted by 5 (cm$^3$/mol) and -5 (cm$^3$/mol) respectively for better visualization.}
	\label{trans_temp}
\end{figure}

From these plots we can identify the antiferromagnetic transition of HoBi at zero field at 5.7 K. When we apply a magnetic field along the [100] direction, we observe that magnetism in HoBi evolves strongly with the magnetic field. We find several field-induced metamagnetic states. Fig.\ref{trans_temp}.a shows the evolution of temperature dependent M/H for several fields between 0.10 T and 5.00 T. For increasing fields the transition is broadened and moved to lower temperatures. The new metamagnetic states emerge for fields higher than 1.50 T. For fields above 4 T, metamagnetic transitions are suppressed. Applying a magnetic field above 5.00 T leads to a saturated paramagnetic state where all the moments are aligned.\\

In order to extract the transition temperatures from the $M/H$ data the temperature derivative of $MT/H$ \cite{Fisher62} is plotted in Fig.\ref{trans_temp}.b. Maxima and shoulders of this derivative were taken as the transition temperatures \cite{Tomy95}. Low field plots show sharp maxima that can be associated to the entrance into the antiferromagnetic state. As the field is increased first a high temperature and then a low temperature shoulder (both associated with the metamagnetic states) emerge and the peak becomes smaller and broader. For higher fields the peak disappears completely and only the two shoulders remain. Fig.\ref{trans_temp}.c shows three representative curves of the series containing
 the different features we associate with the transition temperatures.\\

\begin{figure}[htb!]
	\begin{center}
		\includegraphics[width=0.9\columnwidth,clip]{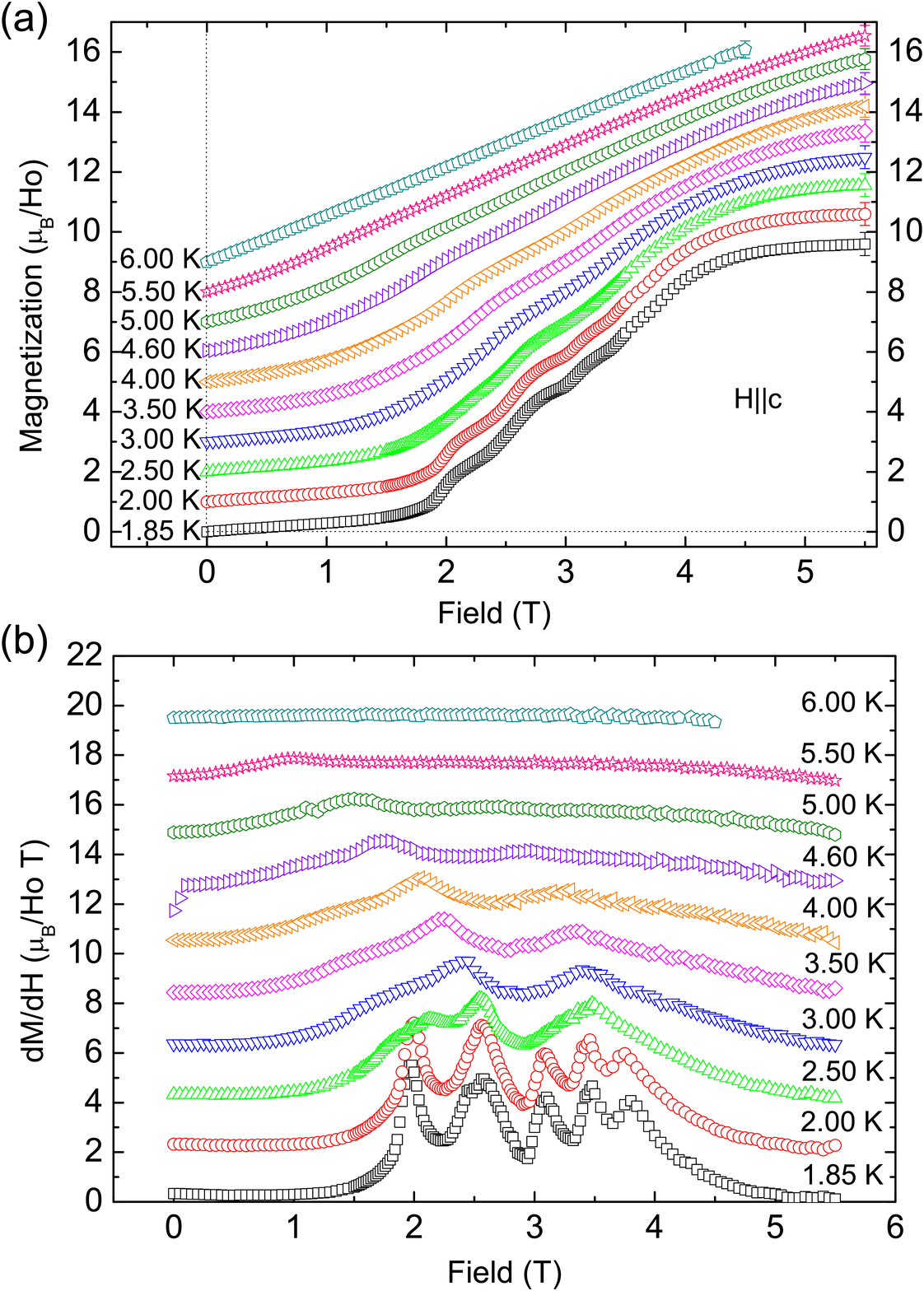}
	\end{center}
	\caption{(a) Magnetization measurements with the data shifted by 1$\mu_B$ between different temperatures. The low temperature data show a cascade of up to five transitions that disappear as the temperature is increased. (b) We plot the derivative of the magnetization with the magnetic field as a function of the magnetic field. Peaks correspond to transitions, providing the phase diagram discussed in the next figure. Derivative data have been shifted by $2 \mu_B/Ho T$ for the different temperatures.}
	\label{trans_field}
\end{figure}

We have also measured magnetization as a function of the magnetic field at different fixed temperatures up to 5.50 T. Previous measurements in Ref. \cite{Ott84} find significant hysteresis in most transitions. Here we find practically no hysteresis. We ascribe this difference to the crystal growth method, which, in our case, should lead to more homogeneous samples. In our measurements, at saturation, magnetization at low temperatures reaches the 10$\mu_B$ expected from Ho$^3+$ ions within the mass error. Fig.\ref{trans_field}.a show the magnetization for temperatures between 1.85 K and 6.00 K (above the zero field transition temperature). The low temperature data show a cascade of field-induced transitions that disappear as the temperature is increased until no transition is seen at 6.00 K.\\ 

Metamagnetic transitions in M(H) data are easily determined from the derivative dM/dH (Fig.\ref{trans_field}b). Five clean transitions can be seen as maxima in 1.85 K and 2.00 K data. For 2.50 K only three maxima can be seen but a broad feature emerges around 2 T and its trace can be followed up to at least 4.00 K. Between 3.00 K and 5.00 K two maxima can be followed, while only one remains at 5.50 K. At 6.00 K magnetization is a straight line and all transitions have disappeared.\\

\section{Conclusions}
	
\begin{figure}[htb!]
	\begin{center}
		\includegraphics[width=0.9\columnwidth]{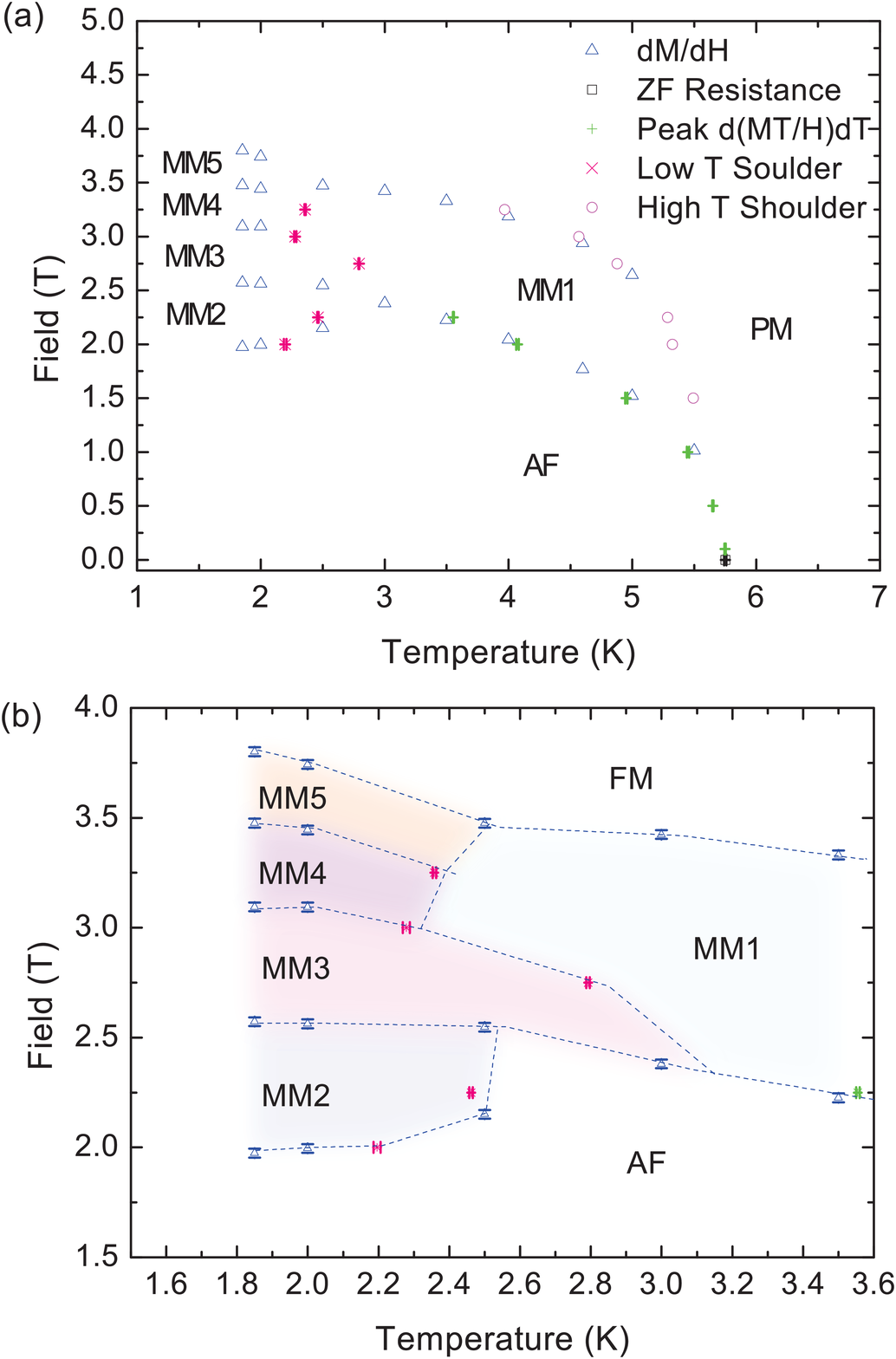}
	\end{center}
	\caption{(a) In the H-T phase diagram several magnetic states can be differentiated. For low fields HoBi is paramagnetic (PM) down to $T_N$ when it enters the antiferromagnetic (AF) phase. When higher external fields are applied up to five metamagnetic (MM) states appear. (b) Detail of the lower temperatures phase diagram. Lines are guides to the eye.}
	\label{phase_diagram}
\end{figure}

With the data from $dM/dH$ and $d(MT/H)/dT$ we obtain a magnetic phase diagram shown in Fig.\ref{phase_diagram}. For the region above 3.00 K three different areas are clearly differentiated: the paramagnetic (PM) state above $T_N$, the antiferromagnetic (AF) below this transition and a metamagnetic state that is labeled as MM1. Below 3.00 K we can differentiate four more metamagnetic states labeled MM2-5.\\

In Fig.\ref{phase_diagram}.b we show a zoom over the low temperature region of the phase diagram. We observe five phases, which emerge at low temperatures from two magnetic phases. 2.5 K seems to be a characteristic temperature, where the magnetic order becomes more complex and field dependent. Fields higher than 5 T force all magnetic moments to point in the [100] direction, leading the system into an induced-ferromagnetic state.\\

Reference \cite{Ott84} studied HoBi single crystals finding a cascade of six metamagnetic transitions at 1.5 K in a squeezed sample. These transitions appear for fields lower than the ones we find here. From Maxwell relation $\left(\left( \partial S/\partial P \right)_T = - \left(\partial V/\partial T \right)_P\right)$, and the positive thermal expansion found in ref. \cite{Ott84}, we deduce that pressure and stress should give lower transition temperatures.\\


To summarize, we have grown and characterized single crystals of HoBi using the excess flux method. Our crystals have a very high RRR, evidencing a low concentration of defects and high electron mean free path. We have built a phase diagram showing a number of stable metamagnetic phases and their evolution with increasing field and temperature. The magnetic phase diagram of HoBi is amazingly rich and deserves further study to understand the spin structure and series of transitions until saturation.\\

\section{Acknowledgments}
We thank discussions with J.V. Alvarez. This work was supported by the Spanish MINECO (Consolider Ingenio Molecular Nanoscience CSD2007-00010 program, FIS2011-23488 and MAT2011-27470-C02-02), by the Comunidad de Madrid through program Nanobiomagnet and by NanoSc-COST program. We also acknowledge Banco Santander. Work at the Ames Laboratory was supported by the US Department of Energy, Basic Energy Sciences, Division of Materials Sciences and Engineering under Contract No. DE-AC02-07CH11358.\\





\section{References}
\bibliographystyle{model1a-num-names}
\bibliography{Revised_Bibliography}{}







\end{document}